\documentclass[12pt]{article}
\usepackage{oldgerm}
\usepackage{yfonts}
\usepackage{euler}
\usepackage{graphics}
\usepackage{bm}
\usepackage{graphicx}
\usepackage{epstopdf}
\usepackage{amsmath}
\usepackage{amssymb}
\usepackage{amstext}
\usepackage{amscd}
\usepackage{amsfonts}
\usepackage{appendix}
\usepackage{color}
\DeclareMathAlphabet{\EuFrak}{U}{euf}{m}{n}
\DeclareMathAlphabet{\EuScript}{U}{eus}{m}{n}

\newcommand{\nd}{\noindent}

\title{{\bf Classical and \textcolor{blue}{Quantum} Field-Theoretical 
approach to the non-linear q-Klein-Gordon Equation }}

\author{{A. Plastino$^1$and M. C. Rocca$^1$}\\
\small{$^1$ La Plata National University and   Argentina's National Research Council}\\
\small{(IFLP-CCT-CONICET)-C. C. 727, 1900 La Plata - Argentina}}

\date{\today}

\begin{document}

\maketitle

\begin{abstract}

\color{blue}{In the wake of efforts made in [EPL {\bf 97}, 41001 (2012)]
and [J. Math. Phys. {\bf 54}, 103302 (2913)], we extend them
here by developing the conventional Lagrangian treatment of
a classical field theory  (FT) to the q-Klein-Gordon
equation advanced in [Phys. Rev. Lett. {\bf 106}, 140601 (2011)] and
[J. Math. Phys. {\bf 54}, 103302 (2913)],
and the quantum theory corresponding to $q=\frac {3} {2}$.
This makes it possible to generate a putative conjecture regarding black matter. 
Our  theory reduces to the  usual  FT for
$q\rightarrow 1$.}

\color{blue}{\nd {\bf Keywords:}Non-linear Klein-Gordon equation;
Classical Field Theory, Quantum Field Theory.\\}
\color{blue}{\nd {\bf PACS:} 11; 11.10.-z; 11.10.Ef; 11.10.Lm; 
02.30.Jr}\normalcolor

\end{abstract}

\newpage

\renewcommand{\theequation}{\arabic{section}.\arabic{equation}}

\setcounter{equation}{0}

\section{Introduction}

\color{blue}Motivated by the need for understanding a number 
of physical phenomena related to complex
systems, interesting proposals for localized solutions have
been proposed in the last five years, based on
modifications of the linear Klein-Gordon and Schrodinger equations. This is done by turning them into nonlinear equations     (NLKG and NLSE, respectively) \cite{tp0,tp1}.  In the wake of efforts made in \cite{tp0,jmp}, we extend them here by developing a {\sf conventional} classical field theory  (FT)
 corresponding to the q-Klein-Gordon equation of \cite{tp1} ({\it the FT of \cite{jmp} is not the customary one, but the higher order FT of \cite{tp10,tp11}}, see below).
 We also advance  the concomitant quantum theory for $q=\frac {3} {2}$.

\nd The NLSE may be employed  for
describing components of dark matter. The structure
of the action variational principle leading to the NLSE
implies  that it might describe particles that do not interact
with the electromagnetic field \cite{tp0}.  Note also that
 the NLSE exhibits a remarkable similarity with
the Schrodinger equation associated to a particle with
a time-position dependent effective mass \cite{40,41,42,43}, involving
 quantum particles in nonlocal potentials (e.g. the energy density functional treatment of the quantum
many-body problem \cite{44}).

\nd We first develop the conventional classical field theory (CFT) associated to the
q-Klein Gordon equation proposed in \cite{tp1}
and deduced in \cite{tp2} from the hypergeometric differential equation  (HDE).
We define the corresponding physical fields via an analogy with treatments in string theory \cite{tp4}
for defining physical states of the bosonic string.
Our ensuing theory reduces to the conventional Klein-Gordon (KG) field theory for $q\rightarrow 1$.
As a second step we develop the quantum theory for
$q=\frac {3} {2}$.

\nd Recently, Rego-Monteiro and Nobre \cite{jmp} advanced an
interesting classical field theory for the generalized 
q-Klein-Gordon equation of \cite{tp1} through the use of
Lagrangian procedures for Higher order equations.
This valuable effort deserves an extension, that will be
tackled here. More to the point:

\begin{itemize}
\item  Rego-Monteiro and Nobre \cite{jmp} use the Higher
Order Lagrangian Procedures of Bollini and Giambiagi \cite{tp10,tp11},
while ours is the usual Lagrangian treatment.

\item They are unable, in their procedure, to throw away
total divergences in the Lagrangian, since, if they do that,
then they do not obtain the correct expression for the
four-momentum of the field, while we do it here.

\item   They do not obtain the physical fields, that is,
the admissible fields for which probability is
conserved.
\end{itemize}

\nd{\bf  Most importantly: we add the Quantum
Field Theory for $q=\frac {3} {2}$, while the
approach of \cite{jmp} is purely classical.}  \normalcolor

\setcounter{equation}{0}

\normalcolor\section{A non-linear q-Klein-Gordon Equation}

\subsection{Classical approach}

Consider then the  q-Klein-Gordon Equation,
advanced in \cite{tp1} y HDE-deduced in \cite{tp2}:
\begin{equation}
\label{eq1.1}
\Box\phi(x_{\mu})+qm^2
\left[\phi(x_{\mu})\right]^{(2q-1)}=0,
\end{equation}
A possible solution to that equation is
\begin{equation}
\label{eq1.2}
\phi(x_{\mu})=[1+i(1-q)(\vec{k}\cdot\vec{x}-\omega t)]^{\frac {1} {1-q}}
\end{equation}
We wish to formulate the CFT associated to (\ref{eq1.1}). We start
with the classical action

\[{\cal S}=\frac {1} {(6q-2)V}\int\limits_M\left\{
\partial_{\mu}\phi(x_{\mu})
\partial^{\mu}\psi(x_{\mu})+
\partial_{\mu}\phi^+(x_{\mu})
\partial^{\mu}\psi^+(x_{\mu})\right.\]
\begin{equation}
\label{eq1.3}
\left.-qm^2\left[
\phi^{(2q-1)}(x_{\mu})\psi(x_{\mu})+
\phi^{+(2q-1)}(x_{\mu})\psi^+(x_{\mu})\right]\right\}d^nx.
\end{equation}
In ${\cal S}$ we detect the appearance of the
de Klein-Gordon field  $\phi$ the auxiliary field  $\psi$. The second arises because
on the non linearity of the q-Klein-Gordon es no-lineal. We recast the action  (\ref{eq1.3})
as
\begin{equation}
\label{eq1.4}
{\cal S}=\int\limits_M{\cal L}(\phi,\psi,\phi^+,\psi^+,
\partial_\mu\phi,\partial_\mu\psi,
\partial_\mu\phi^+,\partial_\mu\psi^+)d^nx,
\end{equation}
where  $M$ stands for  Minkowski's space and
${\cal L}$ is the pertinent Lagrangian.
From the minimum action principle de get the motion equations for the two fields
\begin{equation}
\label{eq1.5}
\frac {\partial{\cal L}} {\partial\phi}-
\partial_\mu\frac {\partial{\cal L}} {\partial(\partial_\mu\phi)}=0
\;\;\;;\;\;\;
\frac {\partial{\cal L}} {\partial\psi}-
\partial_\mu\frac {\partial{\cal L}} {\partial(\partial_\mu\psi)}=0
\end{equation}
The first equation coincides with (\ref{eq1.1}). The auxiliary field equations is
\begin{equation}
\label{eq1.6}
\Box\psi(x_{\mu})+q(2q-1)m^2
\left[\phi(x_{\mu})\right]^{(2q-2)}\psi(x_\mu)=0.
\end{equation}
The solution associated to (\ref{eq1.2}) is
\begin{equation}
\label{eq1.7}
\psi(x_{\mu})=[1+i(1-q)(\vec{k}\cdot\vec{x}-\omega t)]^{\frac {2q-1} {q-1}}
\end{equation}
For  $q\rightarrow 1$, $\psi$ becomes the conjugated of  $\phi$.  \vskip 3mm

\nd We wish to ascertain that the relations between energy and momentum in (\ref{eq1.2}) remain intact
in our formulation. For this we need to evaluate these two field-quantities.  The field's
 Energy-Momentum is

\[{\cal T}_\mu^\nu=
\frac {\partial{\cal L}} {\partial(\partial_\nu\phi)}\partial_\mu\phi+
\frac {\partial{\cal L}} {\partial(\partial_\nu\psi)}\partial_\mu\psi+
\frac {\partial{\cal L}} {\partial(\partial_\nu\phi^+)}\partial_\mu\phi^++\]
\begin{equation}
\label{eq1.8}
\frac {\partial{\cal L}} {\partial(\partial_\nu\psi^+)}\partial_\mu\psi^+-
\delta_\mu^\nu{\cal L}
\end{equation}
Its expression in terms of the two fields becomes
\[{\cal T}_\mu^\nu=\frac {1} {(6q-2)V}[
\partial^\nu\psi\partial_\mu\phi+
\partial^\nu\phi\partial_\mu\psi+
\partial^\nu\psi^+\partial_\mu\phi^++\]
\begin{equation}
\label{eq1.9}
\partial^\nu\phi^+\partial_\mu\psi^+]-
\delta_\mu^\nu{\cal L}
\end{equation}
The four-momentum is
\begin{equation}
\label{eq1.10}
{\cal P}_\mu=\int\limits_V{\cal T}_\mu^0d^{n-1}x,
\end{equation}
where  $V$ is the Euclidian volume. The time-component of the four-momentum is the field energy (up to spatial divergences)
\begin{equation}
\label{eq1.11}
{\cal P}_0=\frac {1} {(6q-2)V}
\int\limits_V
(\partial_0\psi\partial_0\phi+
\partial_0\psi^+\partial_0\phi^+
-\psi\partial_0^2\phi
-\psi^+\partial_0^2\phi^+)d^{n-1}x.
\end{equation}
Using the solutions  (\ref{eq1.2}) y (\ref{eq1.7}) we find for the energy

\begin{equation}
\label{eq1.12}
{\cal P}_0=\frac {1} {(6q-2)V}
\int\limits_V (6q-2)\omega^2d^{n-1}x,
\end{equation}
or
\begin{equation}
\label{eq1.13}
{\cal P}^0={\cal P}_0=\omega^2
\end{equation}
Up to spatial divergences, the field-momentum  is

\begin{equation}
\label{eq1.14}
{\cal P}_j=\frac {1} {(6q-2)V}
\int\limits_V
(\partial_0\psi\partial_j\phi+
\partial_0\psi^+\partial_j\phi^+
-\psi\partial_0\partial_j\phi
-\psi^+\partial_0\partial_j\phi^+)d^{n-1}x.
\end{equation}
Specializing this for the solutions
(\ref{eq1.2}) y (\ref{eq1.7}) one has

\begin{equation}
\label{eq1.15}
{\cal P}_j=-\frac {1} {(6q-2)V}
\int\limits_V (6q-2)\omega k_jd^{n-1}x
\end{equation}
or

\begin{equation}
\label{eq1.16}
{\cal P}^j=-{\cal P}_j=\omega k_j.
\end{equation}
We see that Eqs.  (\ref{eq1.13}) - (\ref{eq1.16})
are proportional to the energy and momentum of the q-exponential wave (\ref{eq1.1}), while the proportionality constant   is the wave energy $\omega$.
 This happens because we did not use a q-exponential divided by $\sqrt{2\omega}$ as is the case with the usual
 Klein-Gordon field when one appeals to waves
$e^{i(\vec{k}\cdot\vec{x}-\omega t)}$
instead of in place of the more common waves
$\frac {e^{i(\vec{k}\cdot\vec{x}-\omega t)}} {\sqrt{2\omega}}$.

The remedy is to choose the constant appearing in the field action as
$\frac {1} {(6q-2)V\omega}$ instead of $\frac {1} {(6q-2)V}$. In such a case the four-momentum becomes
\begin{equation}
\label{eq1.17}
{\cal P}_\mu\rightarrow\frac {{\cal P}_\mu} {\omega},
\end{equation}
and one finds, as expected,

\begin{equation}
\label{eq1.18}
{\cal P}^\mu=(\omega,\vec{k}),
\end{equation}
in complete agreement with the conventional field formulation. Note that,  from
(\ref{eq1.3}), our theory is not gauge invariant save for $q\rightarrow 1$. This entails  that our  fields cannot interact with light. In other words, for $q \ne=1$,
we can have free massive particles of a nonlinear character,
that seem to be incapable to couple with light. This might suggest a mechanism able  to describe the presence of dark matter \cite{tp0}. \vskip 3mm

\nd As for probability conservation, we define the four-current as
\begin{equation}
\label{eq1.19}
{\cal J}_\mu=\frac {i} {4mV}[\psi\partial_\mu\phi-\phi\partial_\mu\psi+
\phi^+\partial_\mu\psi^+-\psi^+\partial_\mu\phi^+].
\end{equation}
Thus, the four-divergence of the four-current does not vanish. It is now

\begin{equation}
\label{eq1.20}
\partial_\mu{\cal J}^\mu={\cal K},
\end{equation}
where  ${\cal K}$ is
\begin{equation}
\label{eq1.21}
{\cal K}=\frac {i} {4mV}q(2q-2)[\psi\phi^{(2q-1)}-\psi^+\phi^{+(2q-1)}.     ]
\end{equation}
Note that  ${\cal K}$ vanishes for $q\rightarrow 1$.

\nd We appeal then to bosonic string's theory \cite{tp4} and define
(in a similar way to that for the definition of physical states)
the {\it physical} fields as those that
make  ${\cal K}$ to vanish.
The waves  (\ref{eq1.2}) y (\ref{eq1.7}) make ${\cal K}$ to vanish.
Also,

\begin{equation}
\label{eq1.22}
{\cal J}^\mu=(\rho,\vec{j}),
\end{equation}
where $\rho$ is

\begin{equation}
\label{eq1.23}
\rho=\frac {i} {4mV}[\psi\partial_t\phi-\phi\partial_t\psi+
\phi^+\partial_t\psi^+-\psi^+\partial_t\phi^+].
\end{equation}
Note that unlike the usual instance,  $\rho$ is not positive-definite and that  $\vec{j}$ is

\begin{equation}
\label{eq1.24}
{\vec j}=-\frac {i} {4mV}[\psi\nabla\phi-\phi\nabla\psi+
\phi^+\nabla\psi^+-\psi^+\nabla\phi^+]
\end{equation}
All quantities defined in this Section become identical to those of the usual KG.CFT for
 $q\rightarrow 1$. \color{blue}{

\setcounter{equation}{0}

\section{Quantum approach for $q=\frac {3} {2}$}

For $q=\frac {3} {2}$ and $m$ small, field quantization can be
performed perturbatively. We write the corresponding
action as:
\[{\cal S}=\int\limits_M\left\{
\partial_{\mu}\phi(x_{\mu})
\partial^{\mu}\psi(x_{\mu})+
\partial_{\mu}\phi^+(x_{\mu})
\partial^{\mu}\psi^+(x_{\mu})\right.\]
\begin{equation}
\label{eq1.25}
\left.-\frac {3} {2}m^2\left[
\phi^2(x_{\mu})\psi(x_{\mu})+
\phi^{+2}(x_{\mu})\psi^+(x_{\mu})\right]\right\}d^4x.
\end{equation}
Now we define i)  the free action ${\cal S}_0$
and ii) that corresponding to the interaction ${\cal S}_I$
as:
\begin{equation}
\label{eq1.26}
{\cal S}_0=\int\limits_M\left[
\partial_{\mu}\phi(x_{\mu})
\partial^{\mu}\psi(x_{\mu})+
\partial_{\mu}\phi^+(x_{\mu})
\partial^{\mu}\psi^+(x_{\mu})\right]d^4x
\end{equation}
\begin{equation}
\label{eq1.27}
{\cal S}_I=-\frac {3} {2}m^2
\int\limits_M\left[
\phi^2(x_{\mu})\psi(x_{\mu})+
\phi^{+2}(x_{\mu})\psi^+(x_{\mu})\right]d^4x.
\end{equation}
The fields in the interaction representation
satisfy the equations of motion for
free fields, corresponding to the action ${\cal S}_0$.
This is to satisfy the usual massless Klein-Gordon equation.
As a consequence, we can cast the fields
$\phi$ and $\psi$ in the form:
\begin{equation}
\label{eq1.28}
\phi(x_{\mu})=\frac {1} {(2\pi)^{\frac {3} {2}}}
\int\limits_{-\infty}^{\infty}\left[\frac {a(\vec{k})} {\sqrt{2\omega}}
e^{ik_{\mu}x_{\mu}}+
\frac {b^+(\vec{k})} {\sqrt{2\omega}}
e^{-ik_{\mu}x_{\mu}}\right]d^3k
\end{equation}
\begin{equation}
\label{eq1.29}
\psi(x_{\mu})=\frac {1} {(2\pi)^{\frac {3} {2}}}
\int\limits_{-\infty}^{\infty}\left[\frac {c(\vec{k})} {\sqrt{2\omega}}
e^{ik_{\mu}x_{\mu}}+
\frac {d^+(\vec{k})} {\sqrt{2\omega}}
e^{-ik_{\mu}x_{\mu}}\right]d^3k
\end{equation}
where $k_0=\omega=|\vec{k}|$
The quantification of these two fields is immediate and
the usual one, given by:
\[[a(\vec{k}),a^+(\vec{k^{'}})]=[b(\vec{k}),b^+(\vec{k^{'}})]=
[c(\vec{k}),c^+(\vec{k^{'}})]=\]
\begin{equation}
\label{eq1.30}
[d(\vec{k}),d^+(\vec{k^{'}})]=
\delta(\vec{k}-\vec{k^{'}})
\end{equation}
The naked propagator corresponding to
both fields is the customary one, and it is just  the Feynman propagator for massless fields:
\begin{equation}
\label{eq1.31}
\Delta_0(k_{\mu})=\frac {i} {k^2+i0}
\end{equation}
where $k^2=k_0^2-\vec{k}^2$
The dressed propagator, which takes into account
the interaction, is given by:
\begin{equation}
\label{eq1.32}
\Delta(k_{\mu})=\frac {i} {k^2+i0-i\Sigma(k_{\mu})}
\end{equation}
where $\Sigma(k_{\mu})$ is the self-energy.

Let us calculate the self-energy for the field
$\phi$ at second order in perturbation theory.
To this order, the  self-energy is composed of two
Feynman diagrams, of which one is null
(this is easily demonstrated using the regularization
of Guelfand for integrals containing powers of
$x$ \cite{tp5}). Therefore, we have for self-energy the
expression:
\begin{equation}
\label{eq1.33}
\Sigma(k_{\mu})=\frac {9m^4} {4}
\frac {i} {k^2+i0}\ast\frac {i} {k^2+i0}
\end{equation}
The convolution of the two Feynman's propagators 
of zero mass is calculated directly using the theory
of convolution of Ultradistributions \cite{tp6}-\cite{tp9}.
Its result is simply:
\begin{equation}
\label{eq1.34}
\frac {i} {k^2+i0}\ast\frac {i} {k^2+i0}=
i\pi^2\ln(k^2+i0)
\end{equation}
The self-energy is then:
\begin{equation}
\label{eq1.35}
\Sigma(k_{\mu})=\frac {9\pi^2m^4i} {4}\ln(k^2+i0)
\end{equation}
As a consequence, the dressed propagator, up to
second order, is given by:
\begin{equation}
\label{eq1.36}
\Delta(k_{\mu})=\frac {4i} {4k^2+9\pi^2m^4\ln(k^2+i0)+i0}
\end{equation}
For both fields $\phi$ and $\psi$ the self-energy
and the dressed propagator  coincide  up to second order.

Note that the current of probability is given by:
\begin{equation}
\label{eq1.37}
{\cal J}_\mu=\frac {i} {4m}[\psi\partial_\mu\phi-\phi\partial_\mu\psi+
\phi^+\partial_\mu\psi^+-\psi^+\partial_\mu\phi^+].
\end{equation}
and it is  verified that:
\begin{equation}
\label{eq1.38}
\partial_\mu{\cal J}^\mu=0
\end{equation}
This implies that the fields defined in the representation
of interaction are physical fields.}

\setcounter{equation}{0}

\normalcolor\section{Conclusions}

\color{blue}{We have here developed further weapons for the formidable 
arsenal being erected
 in the wake of the pioneer work of reference \cite{tp1}, so as to be better able
to face the complex  physics associated  to non-linear quantum equations.

First, we developed the classical field theory  corresponding
to the non-linear
q-Klein-Gordon equation, improving upon the work
of Rego-Monteiro and Nobre \cite{jmp}.

1) Rego-Monteiro and Nobre \cite{jmp} use the Higher
Order Lagrangian Procedures of Bollini and Giambiagi \cite{tp10,tp11},
while we have used the conventional Lagrangian treatment.

2) They were unable, in their procedure, to throw away
total divergences in the Lagrangian, since if they were to do that,
then they would not obtain the correct expression for the
four-momentum of the field. In our procedure we have
removed the total divergences.

3) We have obtained the physical fields, this is,
the admissible fields for which probability is
conserved.

{\bf 4) Most importantly: we have added the Quantum
Field Theory for $q=\frac {3} {2}$,}

We hope that our next stage will be extending things to a quantum field theory
for $q$ a real number such that $1\leq q<2$ a difficult task indeed.}
\normalcolor

\end{document}